\documentstyle[11pt]{report}
\begin{document}
\newtheorem{thm}{Theorem}
\newtheorem{lemma}[thm]{Lemma}
\newtheorem{propo}[thm]{Proposition}
\newtheorem{defin}[thm]{Definition}

\centerline {\Large \bf Homologically Twisted Invariants Related to}

\centerline {\Large \bf (2+1)- and (3+1)-Dimensional State-Sum}

\centerline {\Large \bf Topological Quantum Field Theories }
\bigskip
\medskip

\centerline {\large David N. Yetter}
\smallskip

\centerline {\small \it Department of Mathematics}
\centerline {\small \it Kansas State University}
\centerline {\small \it Manhattan, KS 66506-2602}
\bigskip

\noindent{\small {\bf Abstract:}   We outline a general
construction applicable to the Turaev/Viro [TV], Crane/Yetter [CY]
and generalized
Turaev/Viro invariants (cf. [Y1]) of invariants valued in complex-valued
functions on $H_{D-2}(M^D,Gr_{\cal C})$, where $Gr_{\cal C}$ is the abelian
group of functorial tensor automorphisms on the
artinian tortile category used to construct the TQFT. \bigskip

\noindent{\bf Introduction}
\smallskip

	It is the purpose of this note to introduce a construction
of invariants of 3- and 4-manifolds which combines state-sum
techniques (cf. [TV,CY,Y1,Y2]) with a dependence on (co)homology classes on
the manifold.

	The basic ingredient is an extra piece of structure on the
tensor categories from which the state-sum invariants are constructed:
a (necessarily abelian)
group of functorial tensor automorphisms, which is used as the
coefficient group for homology.

	We assume a familiarity with standard references on
monoidal and tensor categories, e.g. Mac Lane [M], Saavedra-Rivano [S],
Kerler [Ke].

	Throughout all manifolds are assumed to be piece-wise linear
(or, equivalently, smooth).

	A note on terminology:  thoughout, we adopt the convention of using
names originally used by categorists (e.g. monoidal, autonomous, natural
automorphism of the identity functor...) to
refer to not-necessarily abelian categories with a given structure, and
names originally used by algebraic geometers (e.g. tensor, rigid, functorial
automorphism,...) to
refer to abelian categories with a given structure implemented by
(multi)linear functors exact in all variables. We advocate the adoption of
this custom by other authors as a way of cutting down the profusion of
terminology now afflicting quantum topology.

	This work was inspired by unpublished work of Paolo Cotta-Ramusino
at the physical level of
rigor, and by conversations with Cotta-Ramusino
and Louis Kauffman at the XXII International Conference on Differential
Geometric Methods in Theoretical Physics, Ixtapa, Mexico. The results
were actually obtained while the author was in Ixtapa, so he extends
thanks also to the organizers of the conference for their hospitality
and financial support.
\bigskip

{\bf Functorial Tensor Automorphisms}
\smallskip

	As noted in the introduction, the basic ingredient in this
construction is an extra piece of
categorical structure available on the categories from which
the Crane/Yetter [CY] and generalized Turaev/Viro [Y1] invariants were
constructed:

\begin{defin} Let $\cal C$ be a tensor category.
A functorial tensor automorphism of $\cal C$ is a natural isomorphism
$\phi:{\bf 1}_{\cal C}\Longrightarrow {\bf 1}_{\cal C}$, which satisfies
$\phi_{A\otimes B} = \phi_A\otimes \phi_B$ for all objects $A$ and $B$,
and $\phi_I = Id_I$, where $I$ is the identity object for the tensor product.
\end{defin}

	Observe that functorial tensor automorphisms for a subgroup of the
abelian group of natural automorphisms of the identity functor on the
category.

	The categories to which
we will apply this notion are particularly nice,
in that they satsify

\begin{defin} A $k$-linear abelian category $\cal C$ is semisimple if
there exists a set $J$ of objects such that

\begin{enumerate}
	\item Every object is isomorphic to a finite direct sum of objects
in $J$.
	\item The objects of $J$ are simple in the sense that for all
$j \in J$ $Hom(j,j)$ is 1-dimensional over $k$.
	\item For $i,j \in J$ $i \neq j$ $Hom(i,j) = 0$.
\end{enumerate}

	In the case where $\cal C$ is a tensor category, we require that
$I \in J$.
\end{defin}

	For a semisimple tensor category $\cal C$,
functorial tensor automorophisms
have a relatively simple structure.  First, observe that the direct sum
decompositions
required by semisimplicity imply that any natural transformation $\phi $
between
exact functors from  $\cal C$ to any abelian category is completely
determined by its components at objects of $J$, and by the last two conditions
these are determined by a choice of a $k$-scalar $\phi_j$
for each $j \in J$.

	Now for functorial tensor automorphisms, the condition
$\phi_{j\otimes k} = \phi_j \otimes \phi_k$ imposes additional conditions:
whenever $i$ is a direct summand of $j \otimes k$, our choice must
satisfy $\phi_j\cdot \phi_k = \phi_i$.  Using

\begin{defin} A $G$-grading of a semisimple tensor category is a function
$gr:J\rightarrow G$, for $G$ a group, which satisfies $gr(j)gr(k) = gr(i)$
whenever $i$ is a direct summand of $j\otimes k$. The grading group of
$\cal C$, $Gr_{\cal C}$, is the universal group equipped with a $G$-grading
of $\cal C$.  \end{defin}

	The preceding discussion then gives

\begin{propo} \label{grprop} If $\cal C$ is a semisimple tensor category,
there is a canonical isomorphism between the abelian group
of functorial tensor autormorphisms of $\cal C$, and the group of
$k$-characters of $Gr_{\cal C}$.
\end{propo}

	It is left as an exercise to the reader to show that
$Gr_{\cal C}$ exists, and is unique and independent of the choice of $J$ up to
canonical isomorphism.

	 Proposition \ref{grprop} has as a consequence that the group of
functorial automorphisms of $Rep_!(U_q(sl_2))$ is $Z_2$, with the generator
given by $\gamma_\rho = (-1)^\rho$, where following the Temperley/Lieb
conventions [KL], simple objects are indexed by twice the spin
(equivalently by dimension - 1).

	Another observation about functorial tensor automorphisms is worth
mentioning:

\begin{propo} If $H$ is a Hopf algebra (resp. finite dimensional Hopf
algebra), then functorial tensor
automorphisms of $Comod(H)$ (resp. $Mod(H)$) are in canonical
bijection with the set
co-central co-grouplike functionals on $H$ (resp. grouplike elements of $H$).
\end{propo}

	The proof is an elementary exercise in Tannaka-style reconstruction
in the comodule case, which dualizes to give the module case for
finite-dimensional $H$.
\bigskip

\noindent{\bf State-Sum Invariants}
\smallskip

	 Recall that the generalized Turaev/Viro invariant of [Y1] is defined by a
state summation of the form

\[	TV_{\cal C}(M) =
\sum_\lambda N^{-n_0}\prod_{edges e} \Delta_\lambda(e)
\prod_{tetrahedra \sigma} ||\lambda(\sigma )|| \]

\noindent where $\lambda$ ranges over all labellings of a triangulation with
ordered vertices
of edges by a chosen set of simple generating objects (in an artinian
semisimple tortile category $\cal C$) and
of faces by maps chosen from bases for the spaces of maps from the tensor
product of the objects on the ``inbound'' edges of the boundary to the object
on the ``outbound'' edge, where $||\lambda(\sigma)||$ is one of the generalized
6-j symbols of [Y1] (according to the orientation
agreement/disagreement between the ambient orientation and that induced by the
ordering on the vertices), and where $N$ is the sum of the squares of the
quantum
dimensions of the simple objects and $\Delta_\rho$ is the quantum dimension of
the simple object $\rho$.

	Similarly, the Crane-Yetter invariant [CY] in its Temperley-Lieb formulation
(cf. [CKY]) is given by

{\small
\[ CY(W) = N^{n_0-n_1}\sum_{\parbox{.65in}{\tiny
	\begin{center}
labellings $\lambda$
of faces and tetrahedra \end{center} }}
	\prod_{\parbox{.25in}{\tiny
\begin{center} faces $\sigma $ \end{center} }}
\Delta(\lambda(\sigma ))\prod_{\parbox{.5in}{\tiny
\begin{center} tetrahedra $\tau $ \end{center} }}
\frac{\Delta(\lambda(\sigma ))}{ \theta(\lambda(\tau ), \lambda(\tau _0),
\lambda(\tau _2)) \theta(\lambda(\tau ), \lambda(\tau _1),
\lambda(\tau _3))}\prod_{\parbox{.6in}{\tiny
\begin{center} 4-simplexes \end{center} }}\! 15-j \]
}

	To describe the the homologically twisted version of these invariants,
we need to let $A$ be a group of functorial tensor automorphisms on the
underlying tensor category of the theory.  We can then consider homology
and cohomology with coefficients in $A$. Because of the quantum flavor of
our constructions it will be convenient to write (co)cycles, (co)boundaries,
and (co)homology classes multiplicatively.

	Although the construction is given in terms of homology classes, it
is important to note that by Poincar\'{e} duality, we are in fact
giving invariants in both the Turaev/Viro and Crane/Yetter cases which
can be regarded as depending on a 2-dimensional {\em cohomology } class.

	Now, given a 1- (resp. 2-)dimensional homology class $[\alpha]$ with
coefficients in $A$ in the Turaev/Viro (resp. Crane/Yetter) case,
modify the expressions above by replacing
the factors of the form
$\Delta_{\lambda(\phi)}$ for $\phi$ a 1- (resp. 2-)simplex with
$tr(\alpha^\phi_{\lambda(\phi)})$, where $tr$ denotes the internal
trace on the category,
and $\alpha$ is a representative cycle for
$[\alpha]$ supported on the 1- (resp. 2-)simplexes of triangulation,
$\alpha^\phi \in A$ is the coefficient of $\phi$ in $\alpha$ .

	More specifically, we make

\begin{defin}

	The homologically twisted generalized Turaev/Viro invariant of
$(M,[\alpha ])$ for $M$ a 3-manifold, $[\alpha ] \in H_1(M,A)$
is defined by the state-sum

\[	TV_{{\cal C},A}(M,[\alpha ]) =
\sum_\lambda N^{-n_0}\prod_{edges e} tr(\alpha^e_{\lambda(e)})
\prod_{tetrahedra \sigma} ||\lambda(\sigma )|| \]

\noindent where $\cal C$ is a semi-simple tortile category, $A$ is a group of
functorial tensor automorphisms on $\cal C$, and the sum ranges over all
labellings of an ordered triangulation as described above.

	The homologically twisted Crane/Yetter invariant of $(M,[\alpha ])$
for $M$ a 4-manifold, $[\alpha ] \in H_2(M,Z_2)$ is defined by the state-sum

{\small
\begin{eqnarray*}
\lefteqn{CY(W,[\alpha ]) = } \\
 & &	N^{n_0-n_1}\sum_{\parbox{.65in}{\tiny
	\begin{center}
labellings $\lambda$
of faces and tetrahedra \end{center} }}
	\prod_{\parbox{.25in}{\tiny
\begin{center} faces $\sigma $ \end{center} }}
tr(\alpha^\sigma_{\lambda(\sigma )})\prod_{\parbox{.5in}{\tiny
\begin{center} tetrahedra $\tau $ \end{center} }}
\frac{\Delta(\lambda(\sigma ))}{ \theta(\lambda(\tau ), \lambda(\tau _0),
\lambda(\tau _2)) \theta(\lambda(\tau ), \lambda(\tau _1),
\lambda(\tau _3))}\prod_{\parbox{.6in}{\tiny
\begin{center} 4-simplexes \end{center} }}\! 15-j
\end{eqnarray*}
}

\end{defin}

	We now explicitly state and prove in what sense these
state-sums are invariants:

\begin{thm} \label{gtvthm}
	$TV_{{\cal C},A}(M,[\alpha ])$ is independent of the ordered
triangulation used to construct it, and of the the choice of representative
cycle $\alpha $.
\end{thm}

\begin{thm} \label{cythm}
	$CY(M,[\alpha ])$ is independent of the ordered triangulation
used to construct it, and of the choice of representative cycle $\alpha $.
\end{thm}

\noindent{\bf proof of Theorems \ref{gtvthm} and \ref{cythm}:} The proof
proceeds in three stages:  first for a fixed triangulation, we show that
the state-sum is independent of the choice of representative cycle; second
we show that for any instance of the Pachner moves (cf. [P]), we can
change the choice of cycle so that the representing cycle is trivial on
the interior of the region replaced by the Pachner move; and finally we
observe that having shown this, the proof of invariance of the original
untwisted invariant (cf. [Y1], [CY]) under the Pachner moves now carries the
same result for the twisted version. Let $\delta $ denote the dimension
of the homology class used (1 for generalized Turaev/Viro, 2 for Crane/Yetter).

	To see that the state-sum is independent of the choices of
representative cycle (for a fixed triangulation), it suffices to show
that multiplying a cycle by the boundary of a chain supported on a single
$(\delta +1)$-simplex does not change the state-sum. To do this, observe
that we can rewrite that state-sum as a sum indexed by labellings of the
$\delta $-simplexes only of smaller state-sums indexed by the extensions
of the labelling to the $\delta $-simplexes. Specifically, in the
generalized Turaev/Viro case, we have a sum over labellings $\lambda $ of
1-simplexes of sums of the form

\[
\sum_{\parbox{.65in}{\begin{center}\tiny $\mu$ extending $\lambda $ to faces
	\end{center}}}  N^{-n_0}\prod_{edges e} tr(\alpha^e_{\lambda(e)})
\prod_{tetrahedra \sigma} ||\mu(\sigma )|| \]

\noindent while in the Crane/Yetter case, we have a sum over labellings
$\lambda $ of 2-simplexes of sums of the form

{\small
\[ N^{n_0-n_1}\sum_{\parbox{.65in}{\tiny
	\begin{center}
labellings $\mu$
extending $\lambda $ to tetrahedra \end{center} }}
	\prod_{\parbox{.25in}{\tiny
\begin{center} faces $\sigma $ \end{center} }}
tr(\alpha^\sigma_{\lambda(\sigma )})\prod_{\parbox{.5in}{\tiny
\begin{center} tetrahedra $\tau $ \end{center} }}
\frac{\Delta(\lambda(\sigma ))}{ \theta(\mu(\tau ), \lambda(\tau _0),
\lambda(\tau _2)) \theta(\mu(\tau ), \lambda(\tau _1),
\lambda(\tau _3))}\prod_{\parbox{.6in}{\tiny
\begin{center} 4-simplexes $\upsilon $ \end{center} }}\! 15-j(\mu,\upsilon) \]
}

Now, suppose $\alpha = \beta \cdot \partial(\xi^a)$ where $\xi^a$ is the
chain supported on a $\delta $-simplex $\xi $ with coefficient $a$.
We will show in either case that the $\lambda $ sums for $\alpha $ and
$\beta $ are equal.

Now, in either case, $\mu(\xi )$ represents a map in the underlying tensor
category from the tensor product of the labels on the inbound faces of the
boundary to the tensor product of the labels on the outbound faces of the
boundary the cancellation. The cancellation of the components of $a$ and
$a^{-1}$ follows from the monoidal naturality condition on $a$ and
``Schur's lemma''. In the generalized Turaev/Viro case (resp. the
Crane/Yetter case), we move the
occurrences of $a$ and $a^{-1}$ from the loop representing the
trace to the edge in the generalized
6j-symbol (resp. quantum 15j-symbol) corresponding to the same face
(using the graphical Schur's lemma), they then cancel by
monoidal naturality.

That having been shown, our second step follows directly
by excision (and the homology long exact sequence) since the region
removed and replaced by the Pachner moves is contractible.

Finally, the proofs of invariance under the Pachner moves for generalized
Turaev/Viro theory and Crane/Yetter theory involve only the labels interior
to the region removed and its replacement, so provided the support of the cycle
is outside this region, they carry the result for the twisted case as well.
$\Box$

	We can then assemble these invariants into an invariant of the
underlying manifold:

\begin{defin} The total twisted generalized Turaev/Viro invariant (resp.
the total twisted Crane/Yetter invariant) of a 3-manifold (resp. 4-manifold)
is the isomorphism class of the pair

\[ ( H_1(M,A),\ TV_{{\cal C},A}(M,-):H_1(M,A)\rightarrow k \]

\[ \mbox{\rm (resp. }\ \ ( H_2(W,Z_2),\ CY(W,-):\rightarrow {\bf C} )\ ) \]

\noindent where two pairs $(B_i,f_i:B_i\rightarrow k)$ for $i=1,2$ are
isomorphic if there is an group isomorphism $\phi:B_1\rightarrow B_2$
such that $f_1 = f_2(\phi).$
\end{defin}

\noindent {\bf A Surgical Version}
\smallskip

	In the case of a simply connected 4-manifold, we can describe a
related surgical invariant related to the twisted Crane/Yetter invariant
in a way analogous to the relationship between the first invariant of
Broda [B1] and the original Crane/Yetter invariant [CY]. We restrict ourselves
to the simply connected case here to avoid the difficulty of handling
homology classes not represented by spheres.

	Specifically, recall that for a simply connected 4-manifold, every
2-homology class is represented by a product of 2-spheres
(remember we are writing
things multiplicatively instead of additively) with various
coefficients in $A$, and that more specifically the homology group is generated
by chains with non-trivial coefficient only on a single sphere obtained by
taking the core of a 2-handle and attaching the cone on the attaching link
into the original 4-ball.

	We can then give a surgical invariant of the pair $(W,[\alpha ])$
for $[\alpha ] \in H_2(W,Z_2)$ as follows:

	Recall that it is sensible to evaluate link-diagrams with
components labelled by
linear combinations of simple object in $Rep_!(U_q(sl_2))$.
In particular, consider two such linear combinations:

\[ \omega = \sum_{i=0}^{r-2} \Delta_i i \]

\noindent and

\[ \theta = \sum_{i=0}^{r-2} (-1)^i \Delta_i i . \]

\begin{defin}
	Let ${\cal L} = L \cup \dot{L}$ be a surgery description of a
simply-connected 4-manifold as in Kirby [Ki]:  $\cal L$ is a framed link
with two distinguished sublinks---$L$ which for the attaching curves
for a family of 2-handles, and a zero-framed unlink $\dot{L}$ which
represent cores along which ``2-handles will be hollowed out'' (equivalent
to 1-handle attaching data).  Now, choose a family $B$ of components of $L$
such that the chains with non-trivial coefficient on each of the associated
2-spheres form a basis for $H_2(W,Z_2)$. Now, let $< {\cal L} >_{[\alpha ]}$
denote the evaluation of $\cal L$ with those components of $B$ with non-trivial
coefficient in the representation of $[\alpha ]$ labelled $\theta $, and
all other components labelled $\omega $.

	We then define $B(W,-):H_2(W,Z_2)\rightarrow {\bf C}$
by

\[  B(W,[\alpha ]) =
\frac{< {\cal L} >_{[\alpha ]}}{N^\frac{|{\cal L}|+\nu ({\cal L})}{2}} \]

\noindent whenever the Kauffman bracket variable is a principle $4r^{th}$ root
of unity, and by

\[  B(W,[\alpha ]) =
\frac{< {\cal L} >_{[\alpha ]}}
{2^{|\dot{L}|}N^\frac{|{\cal L}|+\nu ({\cal L})}{2}} \]

\noindent whenever the Kauffman bracket variable is a principle $2r^{th}$ root
of unity for $r$ odd, where $N = \sum_{i=0}^{r-2} \Delta_i^2$,
$|\Lambda|$ is the number of components of a link $\Lambda $, and
$\nu({\cal L})$ is the nullity of the linking matrix.
We call the pair $( H_2(W,Z_2), B(W,-):H_2(W,Z_2)\rightarrow {\cal C} )$
the total twisted Broda invariant of $W$.
\end{defin}

\begin{thm}The value of $B(W,[\alpha ])$ is invariant under the Kirby moves
for 4-manifold data and change of the choice of basis for $H_2(W,Z_2)$
and thus $B(W,-):H_2(W,Z_2)\rightarrow {\bf C}$ is an invariant of
$W$.

\end{thm}

\noindent{\bf proof:} Regarding the introduction of attaching data
for cancellable 2-handles, note that a cancellable 2-handle is
never included in the basis, and thus will always be labelled $\omega $, as
will any 1-handle attaching curve. Broda's proof then carries this
part of the result. For handle-sliding, observe that handles labelled
$\omega $ or $\theta $ slide over handles labelled $\omega $
retaining their labels, while they slide over handles labelled $\theta $
and change labels ($\omega $'s become $\theta $'s and vice-versa).

	But this behaviour is precisely the rewriting of the homology
class $[\alpha ]$ under the change of basis corresponding to the
handle-slide, and we are done.

	Alternatively, the following proposition gives a
proof of invariance in terms of the twisted Crane/Yetter invariant.$\Box$

\begin{propo} For any simply connected 4-manifold $W$, for the Kauffman
bracket variable a principle $4r^{th}$ root of unity

\[ B(W,[\alpha ])N^\frac{\chi(W)}{2} = CY(W,[\alpha ]). \]

\end{propo}

\noindent{\bf proof:} A proof identical to that given for the untwisted
invariants (cf. [CKY], [R]) carries this result. In tracing this, the
reader should note one subtlety: we really must extend definition
the twisted
Broda invariant to 2-homology classes represented by sums of
spheres in arbitrary (as opposed to simply connected) 4-manifolds, but
having done this, the proof proceeds as before.$\Box$

\bigskip

\noindent {\bf Conclusion}
\smallskip

	It is the custom in papers on quantum topology to end with
a series of speculations about the implications of the construction/results
obtained. We will forego this custom, save to note that dependence on
cohomology classes is a feature of Donaldson's invariants [D] which was
missing from previous attempts to apply combinatorial methods to
4-dimensional differential topology (cf. [B1,B2,CY]).

	The author is currently computing the total twisted Broda invariants
for Gompf's [G] examples of homeomorphic but non-diffeomorphic 4-manifolds
with simple surgery descriptions.

\vspace{2cm}

\centerline{\large \bf References}
\bigskip

\noindent [B1] Broda, B., {\em Surgical invariants of 4-manifolds}, preprint
(1993).
\smallskip

\noindent [B2] Broda, B., {\em A Surgical invariant of 4-manifolds},
Proceedings of the Conference on Quantum Topology (D.N. Yetter, ed.),
World Scientific, to appear.
\smallskip

\noindent [CKY] Crane, L., Kauffman, L. H. and Yetter, D. N., {\em Evaluating
the Crane-Yetter invariant}, e-preprint hep-th/9309063, and to appear
{\em Quantum Topology} (R. Baadhio and L.H.
Kauffman eds.), World Scientific.
\smallskip

\noindent [CY] Crane, L. and Yetter, D. N., {\em A categorical construction
of 4D topological quantum field theories}, e-preprint
hep-th/9301062, and to appear in {\em Quantum Topology} (R. Baadhio and L.H.
Kauffman eds.), World Scientific.
\smallskip

\noindent [D] Donaldson, S.K., {\em Polynomial Invariants for Smooth
4-Manifolds}, Topology {\bf 29} (1990) 257-315.
\smallskip

\noindent [G] Gompf, R.E., {\em Nuclei of Elliptic Surfaces}, Topology {\bf
30} (1991) 479-511.
\smallskip

\noindent [KL] Kauffman, L. H. and Lins, S. L. {\em Temperley-Lieb Recoupling
Theory and Invariants of 3-Manifolds}, Princeton University Press, to appear.
\smallskip

\noindent [Ke] Kerler, T., {\em Non-Tannakian Categories in Quantum Field
Theory}, in {\em New Symmetry Principles in Quantum Field Theory} (J.
Fr\"{o}lich et al.), Plenum Press (1992).
\smallskip

\noindent [Ki] Kirby, R., {\em The topology of 4-manifolds}, SLNM vol. 1374,
Springer-Verlag (1989).
\smallskip

\noindent [M] Mac Lane, S., {\em Categories for the Working Mathematician},
Springer-Verlag (1971).
\smallskip

\noindent [P] Pachner, U., {\em P.L. Homeomorphic Manifolds are Equivalent
by Elementary Shelling}, Euro. J. Comb. {\bf 12} (1991) 129-145.
\smallskip

\noindent [R] Roberts, J, {\em Skein theory and Turaev-Viro invariants},
preprint (1993).
\smallskip

\noindent [S] Saavedra-Rivano, N., {\em Categories Tannakiennes}, SLNM vol.
265, Springer-Verlag (1972).
\smallskip

\noindent [TV] Turaev, V. and Viro, O., {\em State-Sum Invariants of
3-Manifolds and Quantum 6-J Symbols}, Topology {\bf 31} (1992) 865-902.
\smallskip

\noindent [Y1] Yetter, D.N., {\em State-sum invariants of 3-manifolds
associated to artinian semisimple tortile categories}, Topology and its
App., to appear.
\smallskip

\noindent [Y2] Yetter, D.N., {\em Triangulations and TQFT's}, to appear in
{\em Quantum Topology} (R. Baadhio and L.H. Kauffman eds.), World Scientific.

\end{document}